\documentclass[11pt,a4paper]{article}
\usepackage[textwidth=15cm]{geometry}
\usepackage{authblk}
\usepackage{graphicx}
\usepackage{dcolumn}
\usepackage{amsmath} 
\usepackage{amssymb}   
\usepackage{bm}
\usepackage[small]{caption}
\usepackage{epsf,epsfig,graphicx,color,hyperref}
\usepackage{epsf,epsfig,graphicx,color,hyperref}
\providecommand{\pacs}[1]{\textbf{\textit{pacs---}} #1}
\providecommand{\keywords}[1]{\textbf{\textit{keywords---}} #1}

\begin{document}


\title{Justification for the group-theoretical method as the right way to solve the infinite spherical well in quantum mechanics }

\author{Young-Sea Huang\thanks{yshuang@mail.scu.edu.tw;  yshuang1948@gmail.com}}
\affil{Fl. 6, No. 17, Alley 7, Lane 231, Kung-Kuan Road, Pei-Tou, Taipei 112, Taiwan}

\author{Chun-Hsien Wu\thanks{chunwu@scu.edu.tw}} 
\author{Kung-Te Wu\thanks{A202001@mail.scu.edu.tw}} 
\author{Tzuu-Kang Chyi\thanks{tkchyi@scu.edu.tw}} 
\affil{Department of Physics, Soochow University, Shih-Lin, Taipei 111, Taiwan}

\author{Huitzu Tu\thanks{huitzu2@gate.sinica.edu.tw}}  
\affil{Institute of Physics, Academia Sinica, Taipei 11529, Taiwan}


\maketitle

\begin{abstract}


Recently, the problem of the infinite spherical well was solved by the group-theoretical method to resolve all the peculiarities in the currently accepted solution 
[DOI: 10.13140/RG.2.2.18172.44162 (Researchgate, 2017)]. 
With a view to further justifying the group-theoretical method, the problem is first studied from the viewpoint of classical mechanics. 
Then the radial probability densities predicted by classical mechanics are compared 
with those predicted from solutions of the problem obtained by the group-theoretical method. 
The comparisons clearly indicate the convergence of predictions of quantum  mechanics 
and classical mechanics in the limit of large eigen-energies. 
Therefore, the group-theoretical method is justified as the right way to solve the problem of the infinite spherical well. 

\end{abstract}



\section{Introduction}

Three peculiarities in the currently accepted solution of the problem of the infinite spherical well~\cite{Griffiths,Gasiorowicz} were pointed out  
recently~\cite{yshuang3,yshuang4}. 
In addition, the problem was solved by the group-theoretical method such that all the peculiarities were resolved~\cite{yshuang4}.  
The present work is to further justify the group-theoretical method as the right way 
to solve the problem.
The paper is organized as follows. 
In Sec.~\ref{sec:1}, the problem is studied from the viewpoint of classical mechanics. 
In Sec.~\ref{sec:2}, the problem is solved by group-theoretical method. 
In Sec.~\ref{sec:3}, the radial probability densities predicted by quantum mechanics 
and classical mechanics are compared.  
Concluding remarks are given in Sec.~\ref{sec:4}.

\section{Classical description of motion of a particle inside a spherical box}
\label{sec:1}

\begin{figure}[ht!]
\begin{center}
\includegraphics[width=0.55\textwidth, clip=]{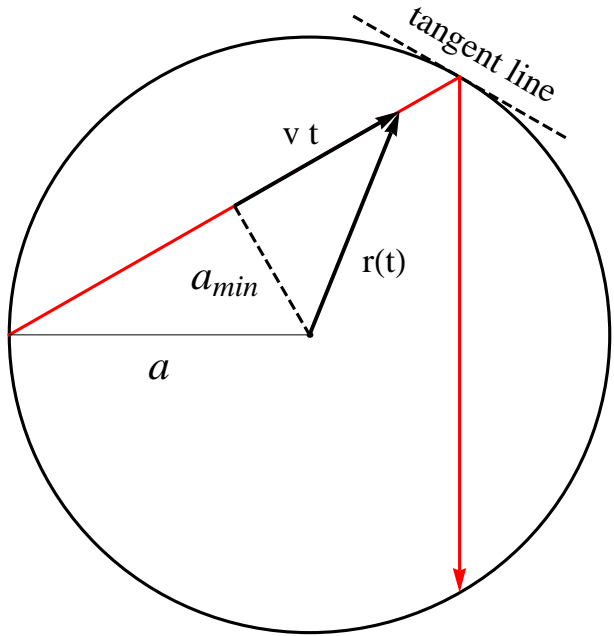}
\end{center}
\caption{\label{fig1} (Color online) Geometry of a trajectory of a particle in a  
two-dimensional circular billiard.}
\end{figure}

Consider a particle of mass $\mu$ moving inside a spherical box of radius $a$. 
The particle moves on one of  planes of the largest circular cross section of the 
sphere. 
Owing to spherical symmetry, it is sufficient to examine the particle moving inside a two-dimensional circular billiard of radius $a$.  
As shown in Fig.~\ref{fig1}, the particle bounces with a specular reflection from the tangent line at the circular boundary, and moves with a constant speed  
$v=\sqrt{2E/\mu}$ in a straight line between consecutive bounces, where $E$ is the kinetic energy of the particle.

By geometry of the trajectory of the particle shown in Fig.~\ref{fig1},  
\begin{equation}
\label{eqr}
   r(t) = \sqrt{(v\, t)^2 + a_{min}^{2}}\, ,
\end{equation}
where $a_{min}$ is the perpendicular distance from the center to the trajectory.
From this equation, we have 
\begin{equation}
\label{eqdt}
   dt =  \sqrt{\frac{\mu}{2E}} \frac{r\, dr}{\sqrt{r^2 - a_{min}^{2}}}\, .
\end{equation}
The radial probability of finding the particle in the range $[r,\, r + dr]$ is directly proportional to the time $dt$ of the particle spent in that range. 
Therefore, the classical radial probability density of the particle is  
\begin{equation}
\label{clasRadProb}
   P_{cl}(r) =  \frac{r}{\sqrt{a^2 - a_{min}^{2}} \sqrt{r^2 - a_{min}^{2}}}\, ,
\end{equation}
where $a_{min} \le r$.
In addition, $a_{min} = L / \sqrt{2 \mu E}$, because the angular momentum of the 
particle relative to the center is $L=\mu v a_{min}$. 
For the case of purely radial motion, $P_{cl}(r)=1/a$ due to $a_{min}=0$  
corresponding to $L=0$. 
For the case of almost purely angular motion, $P_{cl}(r) \rightarrow \delta(r-a)$ as $a_{min}\rightarrow a$.  
For this case, the particle can be found only in the region sufficiently close to the  boundary.  
Using the dimensionless parameter $\sigma \equiv a_{min}/ a$, and measuring $r$ in units of $a$, Eq.~(\ref{clasRadProb}) reduces to
\begin{equation}
\label{clasRadProb2}
  P_{\sigma}(r) =  \left\{  \begin{array}{cl} 
  \frac{r}{\sqrt{1 - \sigma^{2}} \sqrt{r^2 - \sigma^{2}}} &, \,\,  {\rm if}\,\,  r \ge  \sigma \\
    0 &, \,\,  {\rm if}\,\,  r <  \sigma  
 \end{array}
       \right.
\end{equation}
which depends on  $\sigma = L/ \mu v a$, the dimensionless angular 
momentum of the particle.

\begin{figure}[h]
\begin{center}
\includegraphics[width=0.55\textwidth, clip=]{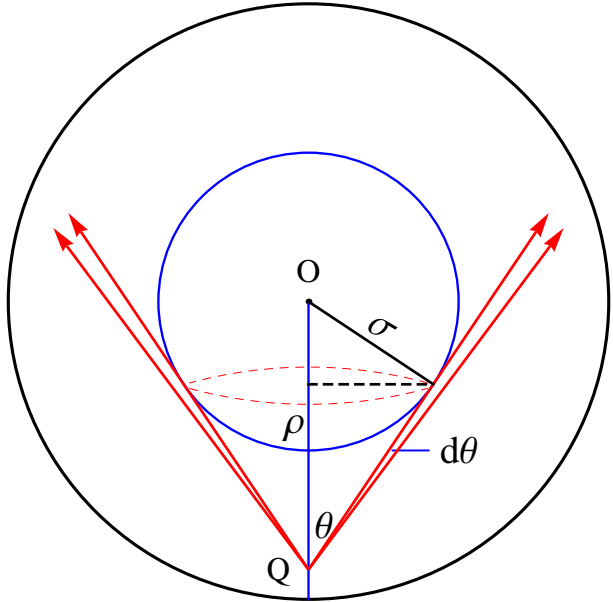}
\end{center}
\caption{\label{fig2} (Color online) As viewed in a plane of the largest circular cross section of the sphere, the radius of the outer circle is one, in units of $a$.  
The distance between the points O and Q is $\rho$, in units of $a$. 
Trajectories of the particle of a dimensionless angular momentum $\sigma$ are 
tangent to the inner circle of the radius $\sigma$.}
\end{figure}

Now, we derive the statistical weight for $P_{\sigma}(r)$. 
The particle has equal probability to move in any direction in the three-dimensional space. 
Thus, the probability for the particle moving in directions confined to a solid angle 
is proportional to the value of that solid angle.  
First, evaluate the probability for the particle having an (dimensionless) angular momentum $\sigma$.
As shown in Fig.~\ref{fig2}, the particle of an angular momentum $\sigma$ moves 
from an arbitrary point $Q$, and its trajectories are tangent to the circle of the 
radius $\sigma$. 
Thus, $\sigma = \rho\, \sin \theta$, where $\rho$ is the distance from the center $O$ 
to the point $Q$. 
The probability for the particle of the angular momentum $\sigma$ moving from the 
point $Q$ is proportional to the solid angle 
$d \Omega(Q,\theta) = 2\pi \sin \theta\,d\theta$.  
Expressing $d \Omega(Q,\theta)$ in terms of the variable $\sigma$, we have 
$d \Omega(Q, \sigma) \propto  \sigma/(\rho \sqrt{\rho^{2}  -  \sigma^{2}})\, d\sigma$. 
In addition, in the three-dimensional space the number of points at the distance 
$\rho$ from the center $O$ is proportional to $ 4\pi \rho^{2}\, d\rho$.
Because the point $Q$ can be at any position between the two concentric spheres, the probability for the particle of the angular momentum $\sigma$ is proportional to
\begin{equation}\label{eqprob}
  d\,\Omega(\sigma) \propto \int_{\sigma}^{1}  \frac{ \sigma \rho}{ \sqrt{\rho^{2}- \sigma^{2}}}\, d\rho \,d\sigma \propto  \sigma   \sqrt{1 - \sigma^{2}}\,d\sigma\, .
\end{equation}

\begin{figure}[h]
\begin{center}\includegraphics[width=0.75\textwidth, clip=]{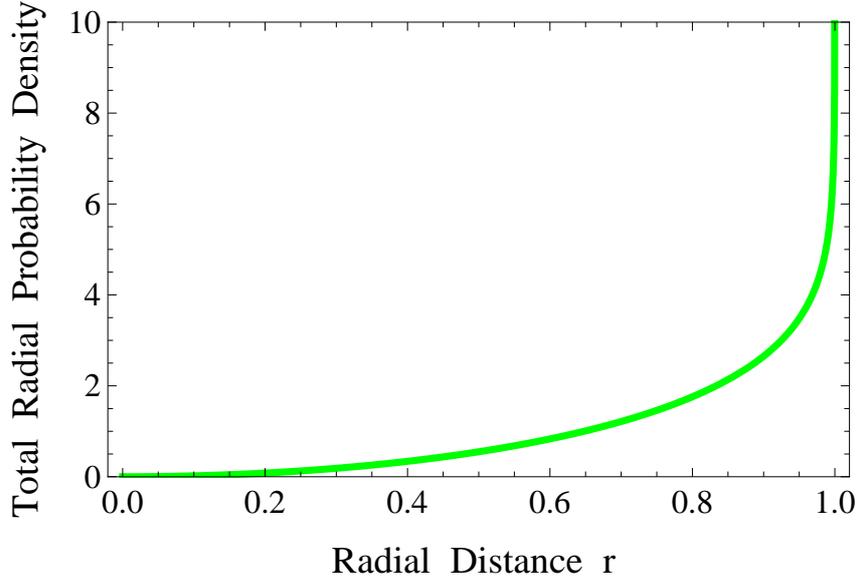}
\end{center}
\caption{\label{fig3} Plot of the classical total radial probability density  
${\bar {\cal P}}_{cl}(r)$.  
The radial distance $r$ is in units of $a$.}
\end{figure}

Second, for the particle of an angular momentum $L=0 \,\, (\sigma =0)$, the time period  between consecutive bounces, corresponding to the radial distance $r$ from $a$ to zero and back to $a$, is $2a/v$.  
For the particle of an angular momentum $L \neq 0\,\, (\sigma \neq 0)$, the time period  between consecutive bounces, corresponding to $r$ from $a$ to $a_{min}$ and back to $a$,  is $2\sqrt{a^{2}- a_{min}^{2}}/v = 2a \sqrt{1- \sigma^{2}}/v $. 
Thus, for one round of consecutive bounces by the particle of $\sigma = 0$, the particle of $\sigma \neq  0$ runs $1 / \sqrt{1- \sigma^{2}}$ rounds of consecutive bounces. Consequently, in the average of the radial probability densities, 
$P_{\sigma}(r)$ need to be weighted by the factor $1 / \sqrt{1- \sigma^{2}}$.   
Then, averaging  $P_{\sigma}(r)$ over all angular momenta with the weight 
$d \Omega(\sigma) / \sqrt{1- \sigma^{2}}$ yields the total radial probability density  
\begin{equation}\label{clasRadProbMean}
  {\bar {\cal P}}_{cl}(r)  = A \int_{0}^{1} P_{\sigma}(r) 
  \frac{d \Omega(\sigma)}{\sqrt{1- \sigma^{2}}}  = r\, 
  ln\left(\frac{1+r}{1-r} \right)\, ,
\end{equation}
where $A$ is the normalization constant. 
The probability ${\bar {\cal P}}_{cl}(r)$ is independent of the energy $E$ of the particle. 
As shown in Fig.~\ref{fig3}, ${\bar {\cal P}}_{cl}(r)$ is increasing as $r$ approaches the boundary.  
It is more likely to find the particle in the region near the boundary than in the vicinity of the center. 
Starting at any position and moving in any direction, the particle will eventually 
hit on the surface of the sphere. 
The probability of the particle hitting on the surface of the sphere is not zero.

\section{The infinite spherical well solved by the group-theoretical method}
\label{sec:2}

Although the infinite spherical well solved by group-theoretical method was already presented in the paper~\cite{yshuang4}, we briefly recapitulate it here for the sake 
of convenient referencing.  
Consider a particle of mass $\mu$ being confined in a well of spherically symmetric potential
\begin{equation}
\label{InfSphWell}
   V(r) = \left\{  \begin{array}{cl} 
               0 &,  \,\,  {\rm if}\,\,  r \le a \\
                \infty   &, \,\, {\rm if}\,\,  r > a 
                \end{array}
       \right.
\end{equation}
The time-independent Schr{\"o}dinger wave equation for the physical system is
\begin{equation}
\label{TIdSchrEq}
  \hat{\rm H}\, \psi({\bf r}) = \left[ - { \hbar^{2} \over 2 \mu } 
  \bigtriangledown^{2}  + \,V( r) \right] \psi({\bf r}) = E\, \psi({\bf r})\, . 
\end{equation}
In terms of the spherical coordinates $(r, \theta, \phi)$, the equation becomes
\begin{equation}\label{TIdSchrEqSphCoor}
  \frac{1}{2\mu \, r^2} \left[- \hbar^{2} \, \frac{\partial}{\partial\, r} 
  ( r^{2} \, \frac{\partial }{\partial\, r} ) + \hat{\rm L}^2 \right]\psi + 
  V(r) \psi  = E\, \psi\, . 
\end{equation}
Here, the operator $\hat{\rm L}^2$ is the square of the angular momentum operator 
$\hat{\bf L}$,
\begin{equation}
\label{SqAngMom}
   \hat{\rm L}^2 = - \hbar^{2} \left[ \frac{1}{ {\rm sin}\, \theta} \, 
   \frac{\partial}{\partial \, \theta } 
   ( {\rm sin}\, \theta \, \frac{\partial }{\partial \, \theta } ) + 
   \frac{1}{  {\rm sin}^{2}\, \theta} \frac{\partial^{2} }{\partial\, 
   \phi^{2}}  \right]\, .
\end{equation}
By the separation of variables, substituting $\psi (r, \theta, \phi) = 
R(r) Y( \theta, \phi)$ into Eq.~(\ref{TIdSchrEqSphCoor}) yields the angular equation  
\begin{equation}
\label{TIdSchrEqSphCoor1}
   \frac{1}{ Y} \, \{ \frac{1}{ {\rm sin}\, \theta} \, \frac{\partial }
  {\partial \,    \theta } ( {\rm sin}\, \theta \, \frac{\partial Y }
  {\partial \,   \theta } ) + \frac{1}{  {\rm sin}^{2}\, \theta} 
  \frac{\partial^{2} Y }{\partial\, \phi^{2}} \} = - l(l+1)\, ,
\end{equation}
and the radial equation
\begin{equation}
\label{TIdSchrEqSphCoor2}
   -  \frac{\hbar^{2}}{2 \mu r^{2}} \,  \frac{d}{d\, r} ( r^{2} \, 
   \frac{d R }{d\, r} ) + \left[ V(r) +  \frac{\hbar^{2} l(l+1)}{2 \mu r^{2}} \right] 
   R = E \, R\, .
\end{equation}

The solutions of Eq.~(\ref{TIdSchrEqSphCoor1}) are spherical harmonics:
\begin{equation}
\label{SphHarFun}
   Y^{m}_{l} (\theta, \phi) = (-1)^{(m + |m|)/2}\, \sqrt{ \frac{2l+1 }{4\pi } 
   \frac{( l- |m| )! }{( l+ |m| )!}  }\, P^{m}_{l} ( {\rm cos} \,\theta )\, 
   e^{ i\, m \phi }\, ,
\end{equation}
where $l=0,1,2,3,\cdot\cdot\cdot$, $m = -l,\cdot\cdot\cdot, l\, $ in integer steps, and $P^{m}_{l} ( {\rm cos}\, \theta )$ are associated Legendre functions. 
These spherical harmonics are common eigenstates of the operators  
$\hat{\rm L}^2$ and $\hat{\rm L}_{z}$, i.e.,
$\hat{\rm L}^2\, Y^{m}_{l} =  l(l+1)\, \hbar^2 \, Y^{m}_{l} $ and $\hat{\rm L}_{z}\, Y^{m}_{l} = m\, \hbar \,  Y^{m}_{l} $. 
They are orthonormal,  i.e.,
\begin{equation}
\label{OrthSphHarFun}
   \int_{\phi = 0}^{2 \pi} \int_{\theta =0}^{\pi} \,\, 
   [Y^{m}_{l}(\theta, \phi)]^{*}[Y^{m'}_{l'}(\theta, \phi)]\, 
   \sin \theta\, d \theta \, d \phi = \delta_{ll'} \delta_{mm'}\, .
\end{equation} 
Outside the well $V(r)=\infty$, so the radial wave function $R(r)=0$ for $r> a$. 
Inside the well $V(r) =0$, so Eq.~(\ref{TIdSchrEqSphCoor2}) becomes
\begin{equation}\label{TIdSchrEqSphCoor2a}
  \frac{d^2 R}{d\, r^2} +  \frac{2}{r} \, \frac{d R }{d\, r}  + 
  \left[k^2 - \frac{l(l+1)}{r^{2}} \right] R = 0\, ,
\end{equation}                         
where $ k=\sqrt{2 \mu E}/ \hbar$ and  $0 \le r \le a$.

Substituting $R(r)=\chi(r)/r$ into Eq.~(\ref{TIdSchrEqSphCoor2}) yields 
\begin{equation}
\label{TIdSchrEqSphCoor2c}
  \hat{\rm H}_{r}  \chi(r) = \left[ - \frac{\hbar^2}{2 \mu} \frac{d^{2}}
  {d\, r^{2}} +    \frac{\hbar^2 l(l+1)}{2 \mu r^{2}} \right] \chi(r) = 
  E\, \chi(r)\, ,
\end{equation}
or,
\begin{equation}
\label{TIdSchrEqSphCoor2ca}
   \frac{d^{2} \chi(r) }{d\, r^{2}} + (\, k^{2} -  
   \frac{ l(l+1)}{ r^{2}} \, ) \, \chi(r) = 0\, .
\end{equation} 
The general solutions of Eq.~(\ref{TIdSchrEqSphCoor2ca}) are $r\, j_{l}(k r)$ and 
$r\, n_{l}(k r)$~\cite{Weber}. 
Nonetheless, for $l \ge 1$, $r\,  n_{l} (k r)$ are divergent at $r=0$, and they are not square-integrable. 
Thus, the solutions $r\, n_{l} (k r)$, except  $r\, n_{0} (k r)$, are not permissible physically. 
The solutions of Eq.~(\ref{TIdSchrEqSphCoor2}) under consideration reduce to  
$ j_{l} (kr)$ and $ n_{0} (kr)$.

It is well known that the essential degeneracy due to symmetry of a physical system can be inferred through group-theoretical consideration~\cite{Tinkham,Joshi}. 
The infinite spherical well is invariant under the rotation-inversion group 
$O(3) = SO(3) \otimes {\cal S}_{2} $. 
Here, $SO(3)$ is the group of symmetry due to rotation in the three-dimensional space, and ${\cal S}_{2}$ is the group of symmetry due to inversion.  
Thus, the Hamiltonian $\hat{\rm H}$ of the system commutes with operators 
$\hat{{\rm L}}^2$, ${\rm \hat L}_z$ and ${\rm{\hat \pi}_i}$, where 
$\hat{{\rm L}}^2$ and ${\rm \hat L}_z$ are angular momentum operators associated with spatial rotation, and ${\rm{\hat \pi}_i}$ is the parity operator associated with 
spatial inversion.   
Therefore, eigenstates of the Hamiltonian can be inferred from common eigenstates of these operators $\hat{{\rm L}}^2$, ${\rm \hat L}_z$ and ${\rm{\hat \pi}_i}$.  

The spherical harmonics $Y^{m}_{l} (\theta, \phi)$ are common eigenstates of the operators $\hat{{\rm L}}^2$ and ${\rm \hat L}_z$. 
Thus, eigenstates of the Hamiltonian are of the form  $R_{l}(r)Y^{m}_{l} (\theta, \phi)$,  where $R_{l}(r)$ are solutions of Eq.~(\ref{TIdSchrEqSphCoor2a}). 
Starting with the case $l=0$, $\sin (k r)$ and $\cos (k r)$ are independent solutions of  Eq.~(\ref{TIdSchrEqSphCoor2ca}). 
By requiring $\sin (k r)$ and $\cos (k r)$ to be orthogonal, we have the 
quantized values of $k$: $k_{n} = n \pi /a$, where $n$ are positive integers. 
Then, $j_{0}(k_{n} r)= \sin (k_{n} r)/ (k_{n} r)$ and 
$n_{0}(k_{n} r)= -\cos (k_{n} r)/ (k_{n} r)$ are orthogonal solutions of Eq.~(\ref{TIdSchrEqSphCoor2a}) with $l=0$. 
Thus, $j_{0}(k_{n} r) Y^{0}_{0} (\theta, \phi)$ and $n_{0}(k_{n} r) Y^{0}_{0} (\theta, \phi)$ are the orthogonal solutions of the Schr{\"o}dinger wave equation  Eq.~(\ref{TIdSchrEq}) with $E={\cal E}_{n}$.

To the eigen-energy ${\cal E}_{n}$, for $l>0$, $j_{l}(k_{n} r)$ are also solutions of  Eq.~(\ref{TIdSchrEqSphCoor2a}) with $k=k_{n}$.  
Therefore for $l>0$,  
\begin{equation}
\label{EigenFunNew}
  \psi_{n l m} (r, \theta, \phi) = R_{n l}(r)\,   
  Y^{m}_{l} (\theta, \phi) = A_{n l}\,  j_{l} (k_{n}\,r ) \, 
  Y^{m}_{l} (\theta, \phi)\, ,
\end{equation} 
are eigenstates of the Schr{\"o}dinger wave equation with $E={\cal E}_{n}$, 
where $A_{n l}$ are normalization constants. 
Furthermore, the quantum number $l$ is constrained by the eigen-energy ${\cal E}_{n}$, 
as shown in the following. 
 
Eq.~(\ref{TIdSchrEqSphCoor2c}) can be viewed as the time-independent 
Schr{\"o}dinger wave equation for a particle in one-dimensional motion under an  
{\it effective} potential energy function $\hbar^2 l(l+1)/2 \mu r^{2}$. 
Because the effective potential energy is only a part of the total energy, 
$\langle \chi |\, \hbar^2 l(l+1)/ 2 \mu r^2 \, | \chi \rangle \le \langle \chi |\, \hat{\rm H}_{r} \, |  \chi \rangle =  E$, for any normalized eigenstate 
$| \chi \rangle$ of eigen-energy $E$. 
Therefore, $ \langle \psi_{n l m} |\, \hbar^2 l(l+1)/ 2 \mu r^2 \, | \psi_{n l m} 
\rangle \le \langle \psi_{n l m} |\, \hat{\rm H} \,| \psi_{n l m} \rangle 
= {\cal E}_{n}$. 
Also, 
\begin{equation}
\label{valuesOfL}   
\begin{array}{cl} 
\langle \psi_{n l m} |\, \hbar^2 l(l+1)/2 \mu r^{2} \, | \psi_{n l m}\rangle & 
  = (\hbar^2 l(l+1)/2 \mu ) \int_{0}^{a} (1/ r^2) R_{n l}(r)^{2} \, r^2 dr   \\
 & \geq  (\hbar^2 l(l+1)/2 \mu a^2) \int_{0}^{a} R_{n l}(r)^{2} \, r^2 dr \\
 & = \hbar^2 l(l+1)/2 \mu a^2\, .
  \end{array}
\end{equation}
Consequently, the values of $l$ are constrained by $l(l+1) \leq (n \pi)^2$.

To the eigen-energy ${\cal E}_{n} $, $| \psi_{n l m}\rangle $ are orthogonal for different values of $l$, because $Y_{l}^{m}$ are orthogonal according to Eq.~(\ref{OrthSphHarFun}). 
These states $| \psi_{n l m}\rangle $ ($l>0$ and $l(l+1) \leq (n \pi)^2$) together with  $j_{0}(k_{n} r) Y^{0}_{0} (\theta, \phi)$ and $n_{0}(k_{n} r) Y^{0}_{0} (\theta, \phi)$ are orthogonal eigenstates of the energy level ${\cal E}_{n} $. 
Therefore, the energy level ${\cal E}_{n}$ is degenerate in the quantum number $l$.  
It should be noted that the eigenstates of the energy level ${\cal E}_{n}$ are not necessarily orthogonal to those of the energy level ${\cal E}_{m}$, where $n \neq m$. This type of degeneracy is different from the "accidental" degeneracy of the energy levels of the Hydrogen atom~\cite{McIntosh}.

\section{An alternative perspective on connection between quantum mechanics and 
classical mechanics}
\label{sec:3}

\begin{figure}[t!]
\begin{center}
\includegraphics[width=0.75\textwidth, clip=]{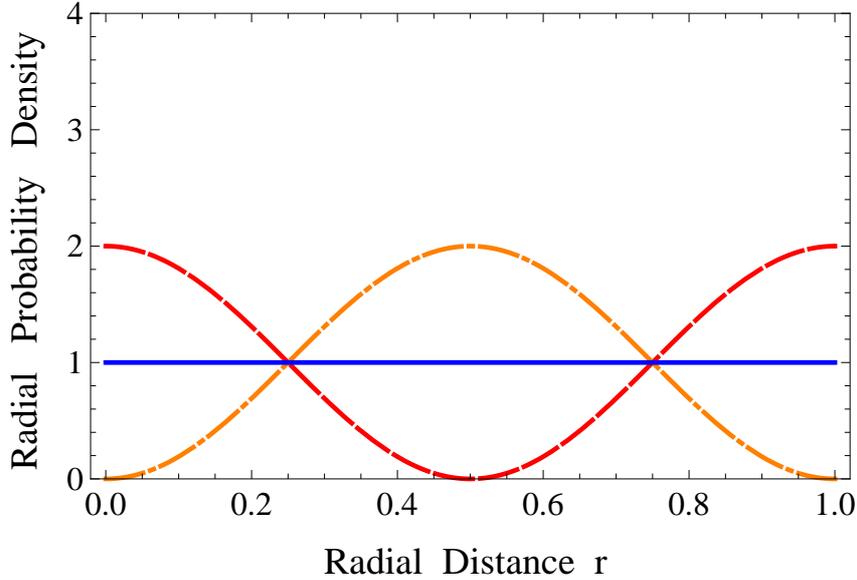}
\end{center}
\caption{\label{fig4} (Color online) For the energy level  ${\cal E}_{1} = \pi^{2} \hbar^{2} / 2 \mu a^{2}$, the radial probability densities of eigenstates 
$\psi_{1 0 0(\rm B)}$ and $\psi_{1 0 0(\rm N)}$, respectively, are shown as a dot-dashed orange line and dashed red line. 
The average of the  radial probability densities of the eigenstates 
$\psi_{1 0 0(\rm B)}$ and $\psi_{1 0 0(\rm N)}$ is shown as a solid blue line. 
The radial distance $r$ is in units of $a$.}
\end{figure}

With a view to establishing connection between quantum and classical descriptions, 
radial probability densities are evaluated from the solutions above, and then compare  with those obtained by classical mechanics in Sec.~\ref{sec:1}.
Consider the first energy level ${\cal E}_{1} =  \pi^{2} \hbar^{2} / 2 \mu a^{2}$. 
Three values of $l$ are allowed: $l= 0,1, 2$, since $ l(l+1) \leq  \pi^2$. 
For $l=0$, there are two orthonormal eigenstates 
$ \psi_{1 0 0(\rm B)}(r, \theta, \phi) = \sqrt{ 2 \pi^2/a^3}\,  j_{0} (\pi\,r /a ) \, Y^{0}_{0}(\theta, \phi)$, and   
$ \psi_{1 0 0(\rm N)}(r, \theta, \phi) = \sqrt{ 2 \pi^2/a^3}\,  n_{0} (\pi\,r /a) \, Y^{0}_{0}(\theta, \phi)$ .
The radial probability densities of the two eigenstates $\psi_{1 0 0(\rm B)}$ and 
$\psi_{1 0 0(\rm N)}$  are, respectively, $P_{10 0(\rm B)}(r) = (2 \pi^2 / a^3)\, |j_{0} (\pi\,r /a)|^2 r^2$ and $P_{10 0(\rm N)}(r) = (2 \pi^2 / a^3)\, |n_{0} (\pi\,r /a)|^2 r^2$. 
Plots of $P_{10 0(\rm B)}(r)$ and $P_{10 0(\rm N)}(r)$ are shown in Fig.~\ref{fig4}. 

The classical counterpart corresponding to the case $l=0$ is the state of the particle 
in radial motion. 
The classical radial probability density of the particle in radial motion is 
$P_{cl}(r)=1/a$.  
The radial probability densities of the two eigenstates are strikingly different from that of the classical counterpart. This implies that each of the two quantum 
eigenstates alone is not analogous to the particle in radial motion.  

The particle have equal probability in any one of the two states, because the two states  have the same eigen-energy.  
Thus, let the mean radial probability density ${\bar P}_{10}(r)$ be the average of, 
with  equal weight, the radial probability densities of the two states.  
The mean of the two radial probability densities ${\bar P}_{10}(r)$ is equal to 
$P_{cl}(r)=1/a$ for the particle in radial motion.

Before proceeding with further comparisons, we detour to discuss about the subtlety of connection between quantum mechanics and classical mechanics. 
Two defects in the conventional comparisons~\cite{Schiff,Townsend,Yoder,Semay}, which illustrate the convergence of predictions of quantum  mechanics and classical mechanics  in the limit of large quantum numbers, were pointed out, but in general have been 
largely ignored~\cite{Leubner}.   
The first defect is that those conventional comparisons try to compare a quantum stationary state with a classical non-stationary state.  
Quantum stationary states are entities intrinsic to quantum mechanics; eigenfunctions 
are not real objects in physical world. 
It seems questionable to expect an {\it abstract} quantum stationary state, such as either $\psi_{1 0 0(\rm B)}$, or $\psi_{1 0 0(\rm N)}$, to be analogous to the 
{\it real} classical state  ---  particle in radial motion.

The second defect is that two conceptually different kinds of probability are involved 
in those conventional comparisons. 
The kind of probability underlying the radial probability density of a quantum 
stationary state, such as $P_{10 0(\rm B)}(r)$, or $P_{10 0(\rm N)}(r)$, 
is {\it quantum} probability that cannot be avoided by any means. 
The kind of probability underlying that of a classical state, such as $ P_{cl}(r)$,  
is {\it classical} probability that can be removed by specifying the initial conditions precisely.  
It is inappropriate to compare a quantum state, either $ \psi_{1 0 0(\rm B)}$, or 
$\psi_{1 0 0(\rm N)}$, with the classical state of the particle in radial motion.

What should be really compared is to compare the classical state with the ensemble of quantum states, $ \psi_{1 0 0(\rm B)}$ and $\psi_{1 0 0(\rm N)}$, as a whole. 
The calculation of the mean radial probability density  ${\bar P}_{10}(r)$ involves the two kinds of probability: first the radial probability densities  
$P_{10 0(\rm B)}(r)$ and $P_{10 0(\rm N)}(r)$ are quantum probability, and then the average over these two radial probability densities, i.e., ${\bar P}_{10}(r)$,  
is classical probability. 
Thus, what should be really compared is to compare 
${\bar P}_{10}(r)$ with $ P_{cl}(r)$.

\begin{figure}[t!]
\begin{center}
\includegraphics[width=0.75\textwidth, clip=]{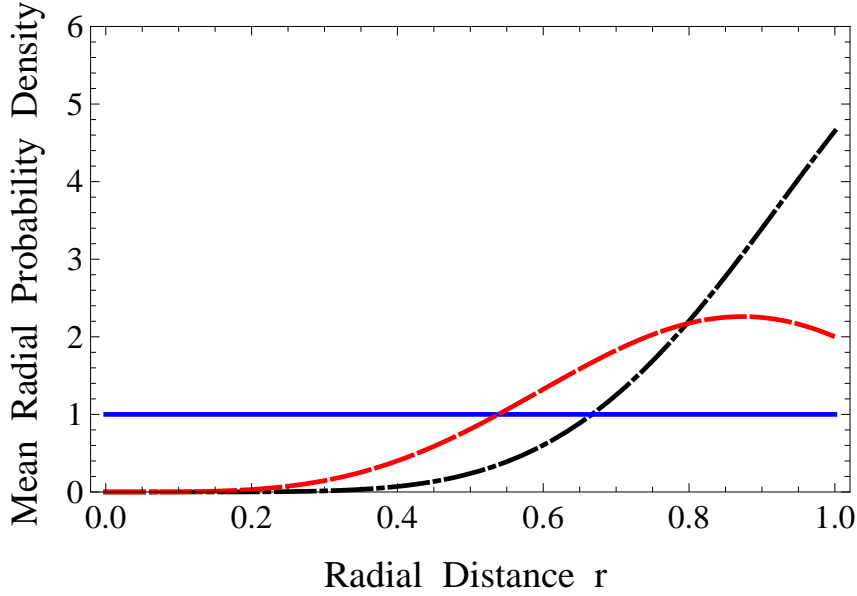}
\end{center}
\caption{\label{fig5} (Color online) For the energy level  ${\cal E}_{1} = \pi^{2} \hbar^{2} / 2 \mu a^{2}$, the mean radial probability densities ${\bar P}_{10}(r)$, ${\bar P}_{11}(r)$ and ${\bar P}_{12}(r)$, respectively, are shown as a solid blue line, dashed red line, and dot-dashed black line.}
\end{figure}

Now, proceed with the cases $l=1$ and $2$.
For $l=1$, the eigenstates are $ \psi_{1 1 m}(r, \theta, \phi) = \sqrt{ 2 \pi^2/a^3}\,  j_{1} (\pi\,r /a) \, Y^{m}_{1}(\theta, \phi)$,
where $m = -1, 0, 1$. 
For $l=2$, the eigenstates are 
$ \psi_{1 2 m}(r, \theta, \phi) = \sqrt{2 \pi^4/ a^3 (\pi^2 -6)}\,  j_{2} (\pi\,r /a) \, Y^{m}_{2}(\theta, \phi)$, where $m = -2, -1, 0, 1, 2$. 
The radial probability densities of the eigenstates  
$\psi_{1 1 m}$ are $P_{11m}(r) = (2 \pi^2 / a^3)\, |j_{1} (\pi\,r /a)|^2 r^2$. 
Because $P_{11m}(r)$ are independent on the values of $m$, the mean radial probability density over the states of $l=1$ is  
${\bar P}_{11}(r) = (2 \pi^2 / a^3)\, |j_{1} (\pi\,r /a)|^2 r^2$. 
Similarly, the radial probability densities of the eigenstates $\psi_{1 2 m}$ are $P_{12m}(r) = (2 \pi^4 / a^3 (\pi^2 -6) )\, |j_{2} (\pi\,r /a)|^2 r^2$. Thus, 
the mean radial probability density over the states of $l=2$ is 
${\bar P}_{12}(r)= (2 \pi^4 / a^3 (\pi^2 -6) )\, |j_{2} (\pi\,r /a)|^2 r^2$. 
The mean radial probability densities of quantum numbers $l= 0,\, 1 \,\, {\rm and} \,\, 2$, i.e., ${\bar P}_{10}(r)$, ${\bar P}_{11}(r)$ and ${\bar P}_{12}(r)$, respectively,  are shown in Fig.~\ref{fig5}.

\begin{figure}[t!]
\begin{center}
\includegraphics[width=0.75\textwidth, clip=]{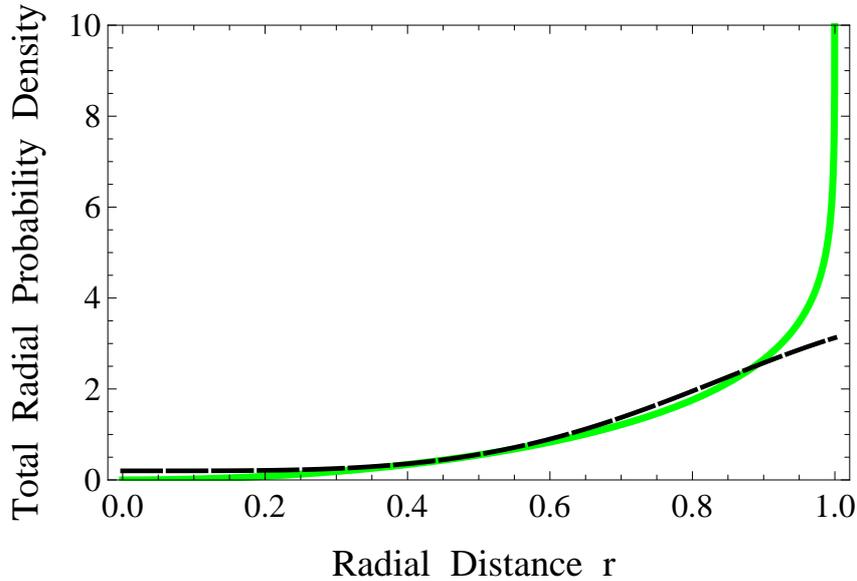}
\end{center}
\caption{\label{fig6} (Color online) The classical total radial probability density  
${\bar {\cal P}}_{cl}(r)$ is shown as a solid green line.  
The total radial probability density  ${\bar {\cal P}}_{1}(r)$ of the particle in the energy level ${\cal E}_{1}$ is  shown as a dashed black line.}
\end{figure}

The total radial probability density ${\bar {\cal P}}_{1}(r)$ of the particle in the energy level ${\cal E}_{1}$  can be calculated as follows. 
For $l=0$, by linear combination of 
$\psi_{1 0 0(\rm B)}$ and $\psi_{1 0 0(\rm N)}$, we have two orthonormal eigenstates 
$\psi_{1 0 0(\rm H1)}(r, \theta, \phi) = \sqrt{  \pi^2/a^3}\,  h^{( 1 )}_{0} (\pi\,r /a) \, Y^{0}_{0}(\theta, \phi)$ and $\psi_{1 0 0(\rm H2)}(r, \theta, \phi) = \sqrt{  \pi^2/a^3}\,  h^{( 2 )}_{0} (\pi\,r /a) \, Y^{0}_{0}(\theta, \phi)$,
where the spherical Hankel functions are 
$h^{( 1 )}_{0}(r) = ( \sin r - i \cos r ) / r = - i\, {\rm Exp}(i\, r) /r$ and  $h^{( 2 )}_{0}(r) = ( \sin r + i \cos r ) / r =  i\, {\rm Exp}(-i\, r) /r$. 
The radial probability densities of $\psi_{1 0 0(\rm H1)}$ and $\psi_{1 0 0(\rm H2)}$ 
are equal to $1/a$, i.e., $P_{10 0(\rm H1)}(r)=1/a$ and $P_{10 0(\rm H2)}(r)=1/a$. 
The general normalized state of the eigen-energy 
${\cal E}_{1}$ is $\psi_{1} = a_{1} \psi_{1 0 0(\rm H1)} + a_{2} \psi_{1 0 0(\rm H2)} + b_{1} \psi_{1 1 1} + b_{2} \psi_{1 1 0} +  b_{3} \psi_{1 1 -1} +  c_{1} \psi_{1 2 2} + c_{2} \psi_{1 2 1} + c_{3} \psi_{1 2 0}+  c_{4} \psi_{1 2 -1} + c_{5} \psi_{1 2 -2} $, with the normalization conditions 
$0\leq \sum_{i=1}^{2} |a_{i}|^2,\, \sum_{j=1}^{3} |b_{j}|^2, \, \sum_{k=1}^{5} |c_{k}|^2  \leq 1$ and $\sum_{i=1}^{2} |a_{i}|^2 + \sum_{j=1}^{3} |b_{j}|^2 +\sum_{k=1}^{5} |c_{k}|^2 =1$. 
Then, the total radial probability density of the particle in  the energy level 
${\cal E}_{1}$ is the average of 
$\sum_{i=1}^{2} |a_{i}|^2 {\bar P}_{10}(r) + \sum_{j=1}^{3} |b_{j}|^2 {\bar P}_{11}(r) + \sum_{k=1}^{5} |c_{k}|^2 {\bar P}_{12}(r)$ over all the possibilities of 
$ a_{i}$, $b_{j}$ and $c_{k}$ satisfying the normalization conditions. 
Therefore, the total radial probability density ${\bar {\cal P}}_{1}(r)$ is just the average of ${\bar P}_{10}(r)$, ${\bar P}_{11}(r)$ and ${\bar P}_{12}(r)$ with the 
weights 2/10, 3/10 and 5/10, respectively. 
The weights on $l$ can be simply argued as due to there are two eigenstates of $l=0$, three eigenstates of $l=1$, and five eigenstates of $l=2$.    
The comparison of ${\bar {\cal P}}_{1}(r)$ and ${\bar {\cal P}}_{cl}(r)$ is shown in Fig.~\ref{fig6}. 
The total radial probability density ${\bar {\cal P}}_{1}(r)$ more or less resembles  
the classical total radial probability density ${\bar {\cal P}}_{cl}(r)$. 

\begin{figure}[h]
\begin{center}
\includegraphics[width=0.75\textwidth, clip=]{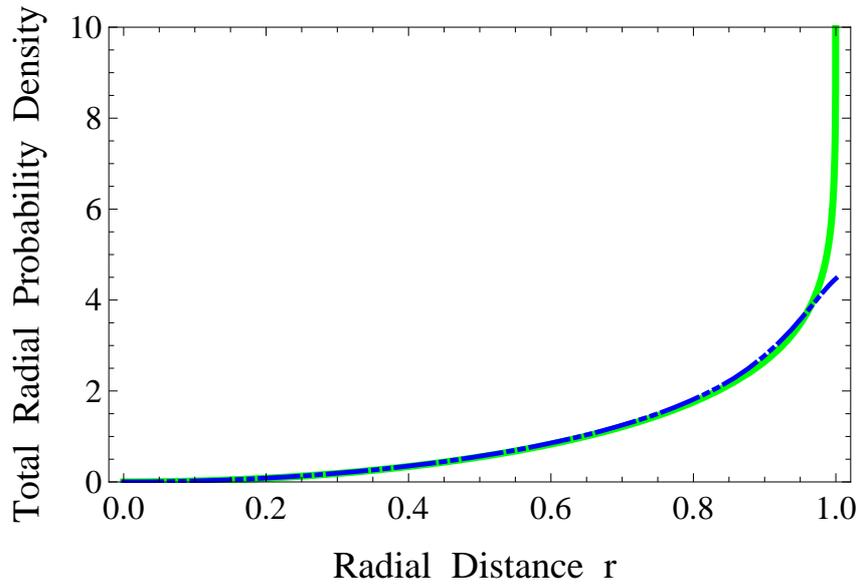}
\end{center}
\caption{\label{fig7} (Color online) The total radial probability density ${\bar {\cal P}}_{10}(r)$ of the particle in the energy level ${\cal E}_{10}$  is shown as a  
two-dots-dashed blue line. 
The classical total radial probability
${\bar {\cal P}}_{cl}(r)$ is shown as a solid green line.}

\end{figure}

\begin{figure}[h!]
\begin{center}
\includegraphics[width=0.75\textwidth, clip=]{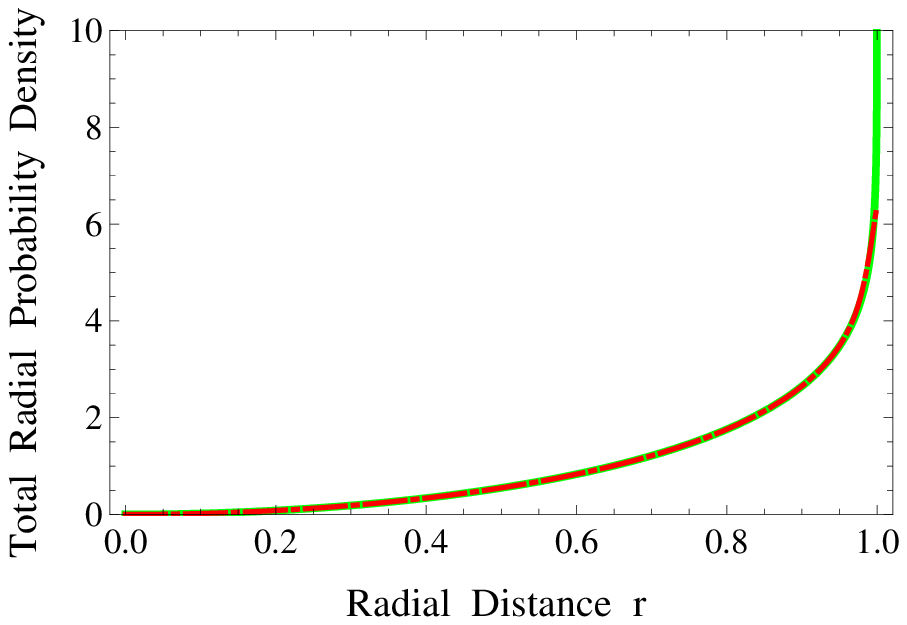}
\end{center}
\caption{\label{fig8} (Color online) The total radial probability density ${\bar {\cal P}}_{100}(r)$ of the particle in the energy level ${\cal E}_{100}$ is shown as a  
dot-dashed red line.  
The classical total radial probability
${\bar {\cal P}}_{cl}(r)$ is shown as a solid green line.}

\end{figure}

\begin{figure}[h!]
\begin{center}
\includegraphics[width=0.75\textwidth, clip=]{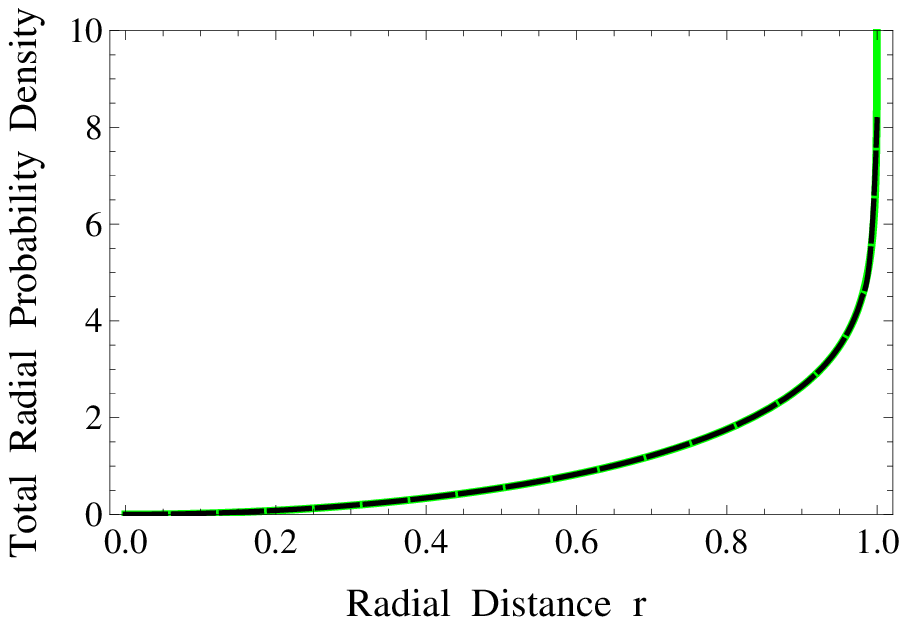}
\end{center}
\caption{\label{fig9} (Color online) The total radial probability density 
${\bar {\cal P}}_{1000}(r)$ of the particle in the energy level ${\cal E}_{1000}$ is shown as a dashed black line.  
The classical total radial probability
${\bar {\cal P}}_{cl}(r)$  is shown as a solid green line.}
\end{figure}

Furthermore, in order to illustrate the convergence of the predictions of quantum mechanics and classical mechanics in the limit of large quantum numbers, 
the total radial probability densities ${\bar {\cal P}}_{10}(r)$, ${\bar {\cal P}}_{100}(r)$, and ${\bar {\cal P}}_{1000}(r)$ of the eigen-energies 
${\cal E}_{10}$, ${\cal E}_{100} $ and ${\cal E}_{1000}$, respectively, are evaluated numerically. 
The  comparisons of them with the classical total radial probability density are 
shown in Fig.~\ref{fig7}, Fig.~\ref{fig8}, and Fig.~\ref{fig9}. 
The comparisons clearly indicate the convergence of the predictions of quantum 
mechanics and classical mechanics in the limit of large eigen-energies.


Both predictions of quantum mechanics and classical mechanics herein show that the particle is more likely to be found in the region closer to the boundary. 
However, according to the prediction of the currently accepted solution, the probability for finding the particle in the region sufficiently close to the boundary is almost vanishing.   
This indicates that the group-theoretical method, rather than the conventional method,  is the right way to solve the problem of the infinite spherical well.

\vfil\newpage
\section{Conclusions}
\label{sec:4}

Quantum mechanics must agree with classical mechanics in an appropriate limit. 
Many conventional comparisons were presented to demonstrate the convergence of predictions of quantum  mechanics and classical mechanics in the limit of large quantum numbers. 
However, those conventional comparisons are highly controversial~\cite{Leubner}. 

The comparisons herein convincingly show the convergence of predictions of quantum  mechanics and classical mechanics in the limit of large quantum numbers. 
Therefore, the group-theoretical method is justified as the right way to solve the infinite spherical well. 
Furthermore, the alternative perspective on connection between quantum mechanics and classical mechanics presented herein may provide insight into the problem of the classical limit of quantum mechanics which has been a subject of heated debate 
since the beginning of quantum 
mechanics~\cite{Leubner,Sen,Castagnino,Makowski,xyhuang,Klein}.


\end{document}